\documentclass[twocolumn,preprintnumbers,showpacs,prb,a4paper,superscriptaddress]{revtex4}

\usepackage{graphicx}
\usepackage{color}
\usepackage{amsmath}
\usepackage{amssymb}
\usepackage{subfigure}
\usepackage{epsfig}

\begin{document}
\title{A precise CNOT gate in the presence of large fabrication induced variations of the exchange interaction strength.}

\author{M. J. Testolin}\email{m.testolin@physics.unimelb.edu.au}\affiliation{Centre for Quantum Computer Technology, School of Physics, The University of Melbourne, Victoria 3010, Australia.}
\author{C. D. Hill}\affiliation{Department of Electrical Engineering and Electronics, University of Liverpool, Brownlow Hill, L69 3GJ, Liverpool, United Kingdom.}
\author{C. J. Wellard}\affiliation{Centre for Quantum Computer Technology, School of Physics, The University of Melbourne, Victoria 3010, Australia.}
\author{L. C. L. Hollenberg}\affiliation{Centre for Quantum Computer Technology, School of Physics, The University of Melbourne, Victoria 3010, Australia.}

\begin{abstract}
We demonstrate how using two-qubit composite rotations a high fidelity controlled-NOT (CNOT) gate can be constructed, even when the strength of the interaction between qubits is not accurately known.  We focus on the exchange interaction oscillation in silicon based solid-state architectures with a Heisenberg Hamiltonian.  This method easily applies to a general two-qubit Hamiltonian.  We show how the robust CNOT gate can achieve a very high fidelity when a single application of the composite rotations is combined with a modest level of Hamiltonian characterisation.  Operating the robust CNOT gate in a suitably characterised system means concatenation of the composite pulse is unnecessary, hence reducing operation time, and ensuring the gate operates below the threshold required for fault-tolerant quantum computation.
\end{abstract}

\pacs{03.67.Lx, 82.56.Jn, 85.30.De}

\maketitle

\section*{Introduction}
%
The ability to correct errors arising from the construction or operation of any quantum computing architecture is essential for a successful implementation.  Without the ability to correct the random and/or systematic errors that arise throughout operation, the implementation of large scale quantum algorithms is hopelessly undermined.  In a realistic device the threshold for fault-tolerant quantum computation is likely to be well below 10$^{-4}$, placing severe constraints on the tolerable magnitude of errors due to decoherence or lack of precision in quantum control.  This work focuses on minimising a particular type of \emph{systematic} error, namely, uncertainty in the coupling strength of two-qubit devices as a result of imperfect fabrication, which causes systematic under- or over-rotations.  We use recently developed two-qubit composite rotations to correct for this uncertainty in the strength of the electron spin exchange interaction in Si:P based architectures\cite{Kane_Nature_393_133_1998, Hollenberg_Phys_Rev_B_74_045311_2006}.  Our results also apply more generally and could be used to correct this type of systematic error in a range of solid-state systems.
%

The strength of the exchange interaction coupling between donors in silicon based solid-state architectures is known to be highly sensitive to donor placement.  The cause of this is the inter-valley interference between the six degenerate conduction band minima of silicon, resulting in oscillations of the exchange coupling strength\cite{Koiller_PRL_88_027903_2002, Wellard_PRB_68_195209_2003, Wellard_J_Phys_Condes_Matter_16_5697_2004, Koiller_PRB_66_115201_2002}.  Exact positioning of donors to better than 2-3 sites is difficult\cite{Schofield_PRL_91_136104_2003} and therefore we expect significant uncertainty in the un-biased strength of the coupling between donors.  The uncertainty in our knowledge of the coupling, leads to error in gate operation.  Systematic errors of this kind are correctable using composite rotations.  Experimental applications already exist in a variety of quantum systems demonstrating the usefulness of composite rotations for ensuring robust operations\cite{Cummins_New_J_Phys_2_1_2000,Collin_PRL_93_157005_2004,Morton_PRL_95_200501_2005,Riebe_Nature_429_734_2004,Schmidt-Kaler_Nature_422_408_2003,Gulde_Nature_421_48_2003}.  Recently two-qubit composite rotations have been considered for systems with uncertainty in their coupling strength\cite{Jones_PRA_67_012317_2003,Hill_quant-ph_0610059_2006}.
%

In this paper, we follow the method for creating a robust controlled-NOT (CNOT) gate developed in Ref.~\onlinecite{Hill_quant-ph_0610059_2006} and quantitatively study the performance of the robust CNOT gate using simulated exchange oscillation data.  We specifically consider the global Si:P electron spin control case where the interaction is of Heisenberg type and gate times are in the $\mathcal{O}$(10-100~ns) regime.  This technique is readily generalisable to any two-qubit Hamiltonian, and for a full treatment, the reader is directed to Ref.~\onlinecite{Hill_quant-ph_0610059_2006}.

Misplacement of donors by only one implantation site can lead to large variations in the exchange coupling strength, even in Si:P systems with voltage bias applied to top gates\cite{Wellard_J_Phys_Condes_Matter_16_5697_2004}, meaning a single application of the composite rotations may not be enough to guarantee a high fidelity CNOT gate.  Concatenating the pulse by feeding it back into itself can help to achieve correction to a higher level, however, performing multiple concatenations costs a large increase in time.  In certain cases using composite rotations alone will not improve the fidelity of the operation above an uncorrected CNOT gate, as the composite rotations are designed to work within a specific uncertainty range.  We show that in unison with Hamiltonian characterisation\cite{Cole_JPhysA_39_14649_2006,Devitt_Phys_Rev_A_73_052317_2006}, the process of experimentally determining a Hamiltonian, a single application of the composite rotations guarantees a high fidelity CNOT operation with an error rate below the fault-tolerant error threshold.  Operating the CNOT gate this way helps remove the need for concatenation, and strikes a balance between fully characterising the system and using composite rotations to construct robust operations.

\section{Constructing robust gates using composite rotations}
%
Composite rotations have been widely used in NMR experiments to correct for pulse length errors and off-resonance effects\cite{Levitt_PRNMS_18_61_1986,Wimperis_JMagnReson_109_221_1994}.  In the case of pulse length errors, a deviation of the field strength from its nominal value leads to systematic under- or over-rotations.  Although originally designed for applications involving single spin quantum systems, composite rotations may be extended to two-spin operations.  In the context of quantum computation, only a certain class of composite rotations, sometimes referred to as fully compensating pulses are applicable, as they work on any initial state.  Using these fully compensating pulses, the application of composite rotations for constructing robust two-qubit gates against pulse length error has already been found for an Ising Hamiltonian\cite{Jones_PRA_67_012317_2003}, and a general two-qubit Hamiltonian\cite{Hill_quant-ph_0610059_2006}.  

\begin{figure}[tb]
\centerline{\includegraphics{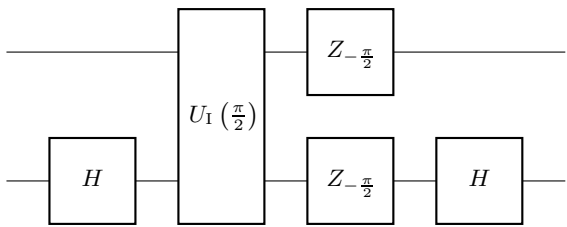}}
\caption{Circuit diagram for a CNOT gate constructed from an Ising interaction, where $H$ is a Hadamard gate, $Z_{-\frac{\pi}{2}}=\exp\left(i\frac{\pi}{4}\sigma_Z\right)$ and $U_{\rm I}\left(\frac{\pi}{2}\right)=\exp\left(-i\frac{\pi}{4}\sigma_Z\otimes\sigma_Z\right)$.}
\label{fig:CNOT}
\end{figure}

In Ref.~\onlinecite{Hill_quant-ph_0610059_2006}, it was noted that for a general two-qubit Hamiltonian expanded in the Pauli basis,
\begin{equation}
H=\sum_{i,j=\{I,X,Y,Z\}}J_{ij}\sigma_i\otimes\sigma_j,
\label{eqn:general_two-qubit_Hamiltonian}
\end{equation}
any interaction term can be effectively extracted using a technique called \emph{term isolation}\cite{Bremner_PRA_69_012313_2004}.  The isolation of a given term will in general not be exact but can be made arbitrarily accurate.  This result is particularly useful and can be used to isolate the Ising coupling term, $J_{ZZ}$, such that we can construct a CNOT gate from this interaction as in Fig.~\ref{fig:CNOT}.
In the case of the Heisenberg interaction with isotropic couplings,
\begin{equation}
H_{\rm H}=J(\sigma_X\otimes\sigma_X+\sigma_Y\otimes\sigma_Y+\sigma_Z\otimes\sigma_Z),
\label{eqn:Heisenberg_Hamiltonian}
\end{equation}
the isolation of the $J_{ZZ}$ term is exact,
\begin{multline}
\exp\left(-iJ_{ZZ}t\sigma_Z\otimes\sigma_Z\right)=-(Z_\pi\otimes I)\exp\left(-iH_{\rm H}t\right)\\
\times (Z_{\pi}\otimes I)\exp\left(-iH_{\rm H}t\right),
\label{eqn:Ising_isolation}
\end{multline}
where for single qubit gates $Z_a$ is a rotation about the $\sigma_Z$ axis by an angle $a$, and similarly for other operators, $J_{ZZ}=2J$, and the global phase factor is included.

\begin{figure*}[tb]
\centerline{\includegraphics[width=17.8cm]{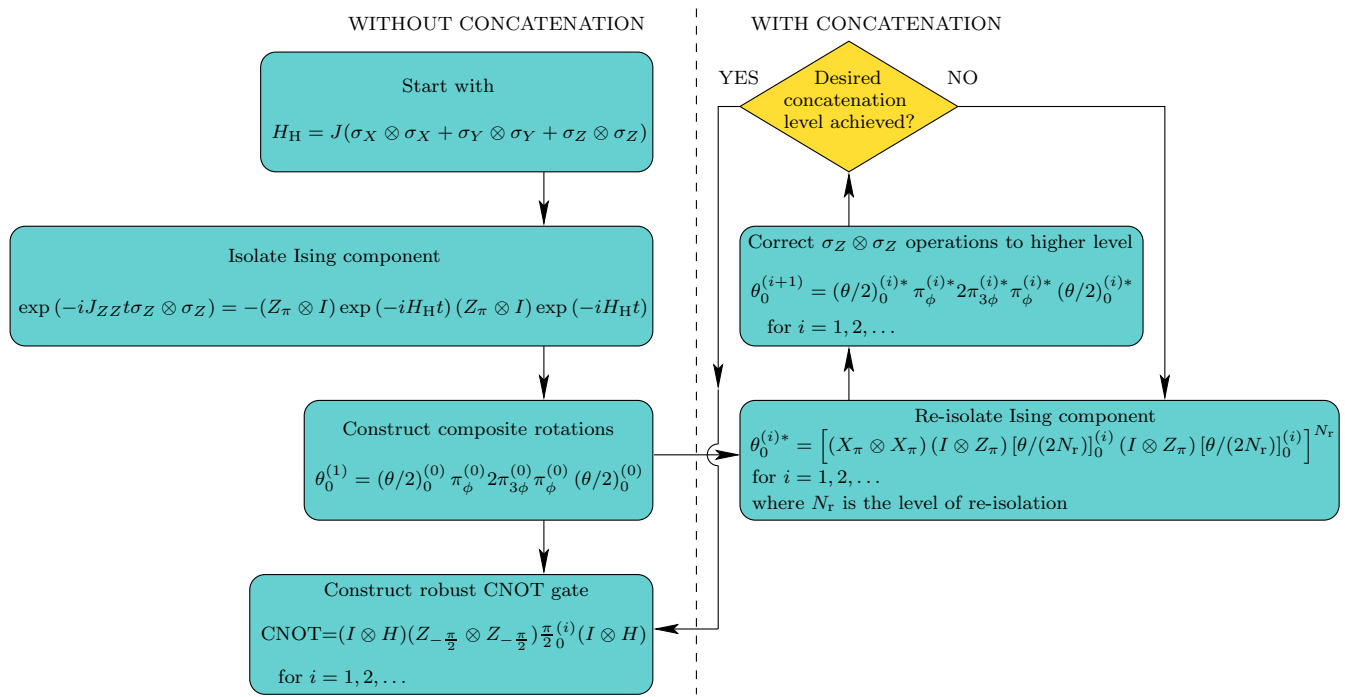}}
\caption{Procedural flowchart for constructing a robust CNOT gate using composite rotations, and concatenating to higher implementation levels.}
\label{fig:procedural_flowchart}
\end{figure*}

We now consider constructing a robust CNOT gate using composite rotations, whereby we replace the interaction term with one created using composite rotations.  Doing this compensates for any uncertainty in our knowledge of the exchange interaction coupling strength, $J$.  In Fig.~\ref{fig:procedural_flowchart} the entire process of constructing a robust CNOT gate from composite rotations is demonstrated schematically.

In an ideal system with a perfectly characterised coupling strength, the evolution operator generated by the Ising interaction is
\begin{equation}
\theta_0\equiv U_{\rm I}\left(\theta\right)=\exp\left(-i\frac{\theta}{2}\sigma_Z\otimes\sigma_Z\right).
\label{eqn:ideal_rotation}
\end{equation}
Here, $\theta_0$ is a two-qubit rotation by an angle $\theta$ about the $\sigma_Z\otimes\sigma_Z$ axis.  In general, $\theta_a$ is a two-qubit rotation by an angle $\theta$ around an axis tilted from the $\sigma_Z\otimes\sigma_Z$ axis towards the $\sigma_Z\otimes\sigma_X$ axis by an angle $a$,
\begin{equation}
\theta_a=\exp\left[-i\frac{\theta}{2}\left(\sigma_Z\otimes\sigma_Z\cos a+\sigma_Z\otimes\sigma_X\sin a\right)\right].
\label{eqn:general_ideal_rotation}
\end{equation}
This two-qubit rotation is achievable via,
\begin{equation}
\theta_a=\left(I\otimes Y_a\right)\theta_0\left(I\otimes Y_{-a}\right).
\label{eqn:off_axis_rotations}
\end{equation}
We make the assumption that all single qubit unitaries are error free, but note that single qubit operations may also be made robust using existing techniques developed in the context of NMR.

In reality a fractional error, $\Delta$, in the two-qubit operation will be present due to the uncertainty in our knowledge of the actual coupling strength, $J_{ZZ}$,
\begin{equation}
\Delta=\frac{J_{ZZ}}{J_P}-1.
\label{eqn:delta_definition_J_ZZ}
\end{equation}
Here, $J_P$ is our prediction of the Ising coupling strength based on the targeted donor separation.  Therefore the actual rotation performed will be
\begin{equation}
\theta_0^{(0)}\equiv U\left(\theta\right)=\exp\left[-i\frac{\theta}{2}(1+\Delta)\sigma_Z\otimes\sigma_Z\right].
\label{eqn:implemented_rotation}
\end{equation}
The superscript of $\theta_a^{(b)}$ in the above equation indicates the \emph{implementation level} of the actual (non-ideal) rotation, with ``(0)'' being an uncorrected implementation and higher levels signifying subsequent corrections from composite rotations.  The \emph{implementation level} should not be confused with \emph{concatenation level}, (e.g., 2$^{\rm nd}$ implementation level is the 1$^{\rm st}$ concatenation level).

\begin{figure}[tb]
\centerline{\includegraphics{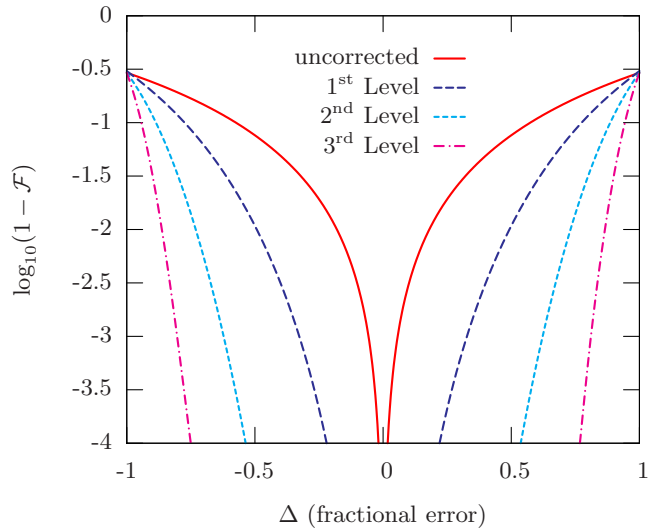}}
\caption{CNOT error, ($1-{\cal F}$), as a function of the fractional error in our knowledge of the coupling strength, $\Delta$, for various implementation levels.  These composite rotations provide improvement over an uncorrected implementation for $\Delta\in (-1,1)$.  The fidelity of a CNOT gate constructed from the Heisenberg interaction using composite rotations was originally calculated in Ref.~\onlinecite{Hill_quant-ph_0610059_2006}.}
\label{fig:error_composite_concatenation}
\end{figure}

It has been previously noted that single qubit composite rotations can be extended to two-qubit composite rotations for use in quantum computation\cite{Jones_PRA_67_012317_2003, Hill_quant-ph_0610059_2006} using fully compensating pulses. A class of these composite rotations known as BB1\cite{Wimperis_JMagnReson_109_221_1994, Cummins_PRA_67_042308_2003} is particularly useful for applications involving quantum computation\cite{Xiao_PRA_73_032334_2006}.  Replacing the pulse $\theta_0^{(0)}$ with the symmetrised BB1 class composite pulse
\begin{equation}
\theta_0^{(1)}=\left(\theta/2\right)_0^{(0)}\pi_{\phi}^{(0)} 2\pi_{3 \phi}^{(0)} \pi_{\phi}^{(0)}\left(\theta/2\right)_0^{(0)},
\label{eqn:BB1_composite_pulse}
\end{equation}
where $\phi=\arccos(-\theta/4\pi)$, will result in a higher fidelity operation.  The fidelity of an operation is defined as
\begin{equation}
{\cal F}=\frac{\left|\mbox{Tr}\left[U^\dagger\left(\theta\right)U_{\rm I}\left(\theta\right)\right]\right|}{\mbox{Tr}\left[U^\dagger_{\rm I}\left(\theta\right) U_{\rm I}\left(\theta\right)\right]}.
\label{eqn:fidelity}
\end{equation}
We may re-isolate the Ising component $J_{ZZ}$ again to arbitrary accuracy as in Fig.~\ref{fig:procedural_flowchart}.  The re-isolated Ising component can then be used to correct to even higher order by passing this pulse back into each of the constituents of Eq.~\ref{eqn:BB1_composite_pulse}, (see Fig.~\ref{fig:procedural_flowchart}).  In principle there is no limit to how often this concatenation can be done, however, the increase in gate time means that in practice this process will be limited by the decoherence time of the system in which the CNOT gate is being implemented.  In Fig.~\ref{fig:error_composite_concatenation} the performance of the uncorrected CNOT gate is compared to the robust gate for various implementation levels, as originally calculated in Ref.~\onlinecite{Hill_quant-ph_0610059_2006}.
Notice each subsequent implementation level performs better over a larger range of the fractional error, $\Delta$.

We now apply the robust CNOT gate to the Si:P architecture with large fabrication induced variations (and hence uncertainty) in the exchange interaction strength.

\section{Correcting for an unknown exchange interaction strength}
%
Systematic errors arising from imperfections in the fabrication process are correctable.  In Kane type architectures\cite{Kane_Nature_393_133_1998,Hollenberg_Phys_Rev_B_74_045311_2006} where phosphorus donors are implanted into an isotopically pure $^{28}$Si matrix, two fabrication processes are being pursued concurrently\cite{Dzurak_QIC_1_82_2001}.  The \emph{top down} approach uses ion beam implantation of phosphorus ions incident on the silicon substrate.  Precise placement of phosphorus donors is limited in this approach due to scattering off the silicon atoms, in a process known as \emph{straggling}.  State of the art top down fabrication results in placement uncertainties of $\mathcal{O}$(10~nm)\cite{Jamieson_APL_86_202101_2005}.  The \emph{bottom up} approach offers atomically precise fabrication using a phosphine gas.  The gas is applied to a hydrogen terminated silicon substrate, where scanning tunneling microscopy has removed individual hydrogen atoms from the hydrogen mono-layer at the desired implantation sites.  Once the phosphorus is integrated into the substrate, the mono-layer is removed and overgrown with silicon.  Small deviations from target implantation by of $\mathcal{O}$(1~nm) (approx. 2-3 sites) can still occur during the annealing process\cite{Schofield_PRL_91_136104_2003}.

\begin{figure}[tb]
\centerline{\includegraphics{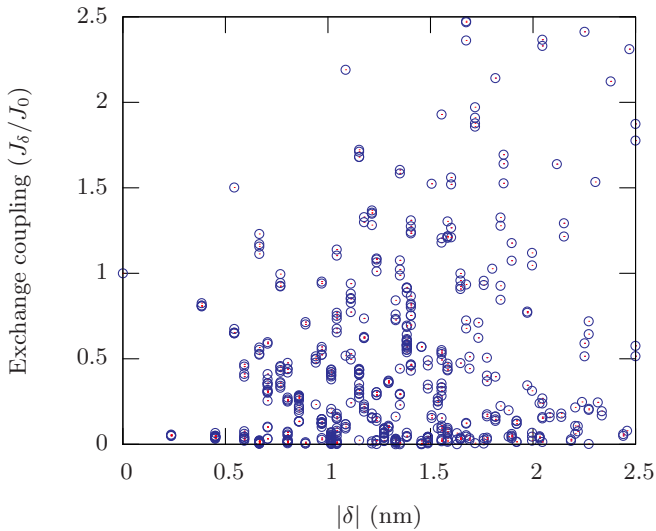}}
\caption{Exchange couplings for donors at fcc lattice sites misplaced by a distance $\delta$ in all directions from the target separation of 20.634~nm (in the [100] direction).  The exchange coupling strengths are given as a fraction of the target coupling strength, $J_0$.}
\label{fig:exchange_coupling_variation}
\end{figure}

The exchange coupling $J$ of the Heisenberg Hamiltonian (see Eq.~\ref{eqn:Heisenberg_Hamiltonian}), is highly sensitive to donor electron wave function overlap.  This means that even small deviations from the targeted implantation sites can lead to large variations in the exchange coupling between donors\cite{Koiller_PRL_88_027903_2002, Wellard_PRB_68_195209_2003}.  Calculated variations in the strength of $J$ for small deviations from the targeted donor separation are shown in Fig.~\ref{fig:exchange_coupling_variation}.  This calculation was performed using the Heitler-London formalism, where the wave functions for the phosphorus donors in silicon were expressed in Kohn-Luttinger effective mass form, with Bloch states explicitly computed using the pseudopotential fit to the band structure.  Details can be found in Ref.~\onlinecite{Wellard_PRB_68_195209_2003}.  Importantly, this type of systematic error is correctable using the composite rotations described above.

We consider donors to be implanted along the [100] lattice direction, as oscillations are suppressed in this direction and the coupling is stronger relative to the [110] and [111] directions, meaning this is the preferred direction for device fabrication.  In an uncharacterised system we assume that the exchange interaction strength is $J_0$ and will be determined by the target donor separation and bias on the control gates.  Fabrication induced donor misplacement will cause the true exchange interaction strength, $J$, to be quite different from $J_0$.  The fractional error in our knowledge of the coupling strength is
\begin{equation}
\Delta_0=\frac{J}{J_0}-1.
\label{eqn:delta_definition_J}
\end{equation}
These composite rotations will only provide an improvement over an uncorrected implementation for $|\Delta_0|<1$.  For $|\Delta_0|>1$ these composite rotations are actually outperformed by the uncorrected implementation, so if $J>2J_0$ then the composite rotations provide a less robust operation.  Interestingly, provided $J\neq 0$, in which case we have no entangling operation, composite rotations will correct for any $J\in (0,2J_0)$.

\begin{figure}[tb]
\centerline{\includegraphics{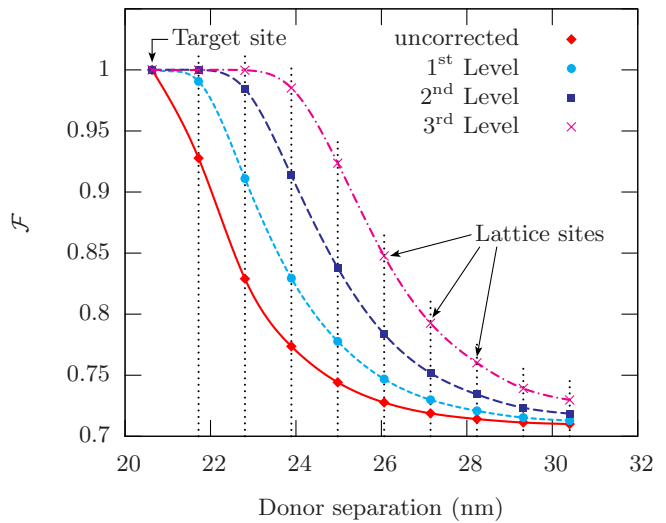}}
\caption{CNOT fidelity as a function of donor separation in the [100] direction for various implementation levels.  The resulting fidelities are determined based on a target donor separation of 20.634~nm.  Note that interpolating curves between lattice sites indicate donor separation scenarios for a given implementation, and vertical dotted lines guide the eye between implementations.}
\label{fig:fidelity_exchange_100_direction}
\end{figure}

Implementing the gate based on the target coupling strength $J_0$, the fidelity of the resulting CNOT operation will be determined by the size of the fractional error $\Delta_0$ in the actual coupling strength.  In Fig.~\ref{fig:fidelity_exchange_100_direction} we demonstrate the resulting CNOT fidelity for a number of donor separations in the [100] direction when the target separation is 20.634~nm.
The results show that using composite rotations improves the fidelity of operation for the CNOT gate.  For example, if the actual separation is 21.720~nm, one application of the composite pulsing scheme improves the fidelity from $\sim 0.93$ to $\sim 0.99$, whilst a second application brings the fidelity above 0.9999.  The successive improvements due to the various levels of pulse concatenation do however come at the expense of operation time.  We examine this issue in the following sections.

\subsection{Gate count}
%
The robust CNOT gate outperforms the uncorrected CNOT gate given an error in the targeted coupling strength, $J_0$, for $|\Delta_0|<1$.  Each level of concatenation provides further improvement, however the cost of this improvement is an exponential increase in the total number of gates required.  An unavoidable consequence of this is an increase in the time required to perform these robust operations.  To be of use for quantum computation we need to be able to perform many precise operations within the decoherence time of the system.  Minimising the time taken to perform a robust CNOT gate is essential.  We show how this can be achieved in Sec.~\ref{sect:Charactersiation} utilising Hamiltonian characterisation.  Below, we consider the actual time costs of concatenated composite pulse correction.

An uncorrected CNOT gate requires only 6 single qubit gates and 2 two-qubit gates.  In comparison, a raw gate count for the number of single qubit gates required in constructing the robust Ising interaction for the CNOT gate yields
\begin{eqnarray}
n_1&=&16,\nonumber\\
n_i&=&10N_{\rm r}(n_{i-1}+2)+6,\quad i=2,3,\ldots~,
\label{eqn:ising_single_qubit_gate_count}
\end{eqnarray}
 where $n_i$ is the number of single qubit gates required for the $i^{\rm th}$ implementation level, and $N_{\rm r}$, which we assume to be constant, quantifies how much we re-isolate the Ising term for pulse concatenation.  Constructing a robust CNOT gate requires an additional 4 single qubit gates, such that the total number of single qubit gates required, $n_i^{\rm 1q}$, is
\begin{equation}
n_i^{\rm 1q}=n_i+4,\quad i=1,2,\ldots~.
\label{eqn:single_qubit_gate_count}
\end{equation}
 The total number of two-qubit gates needed in the robust CNOT construction is
\begin{equation}
n_i^{\rm 2q}=10^iN_{\rm r}^{i-1},\quad i=1,2,\ldots~,
\label{eqn:two_qubit_gate_count}
\end{equation}
again assuming the same $N_{\rm r}$ for each level of concatenation.  Additional re-isolating and concatenation increases the number of single and two-qubit operations required.  We may be able to reduce the total number of single qubit operations by compounding gates however this is not possible for the two-qubit operations.  The process of re-isolating the Ising component of the two-qubit operation slices the rotation into many smaller rotations, as well as adding extra operations.  The limit to how finely we can slice will be decided by the strength of the exchange coupling.  To perform small two-qubit rotations we may require a small $J$ such that the operations evolve slowly enough to be within the realms of experimental pulse timing control.  The viability of using multiple concatenation for constructing robust two-qubit gates lies in tenuous balance between the ability to perform the large number of operations required quickly, and adequate pulse timing control over the small two-qubit rotations.  The strength of the exchange coupling of our system will determine whether these conditions can be satisfied.

\subsection{Gate time}
Each level of concatenation increases the time taken for the robust CNOT operation significantly.  In a working quantum computer this may be problematic as the decoherence time of the system sets an upper limit on how long operations may take.  For phosphorus donors in Si the coherence time, $T_2$, of donor electron spins has been measured to be $T_2>60\ \mbox{ms}$ at 7~K\cite{Tyryshkin_Phys_Rev_B_68_193207_2003}.  We calculate the total time taken for the robust CNOT gate for various implementation levels based on gate times using global control methods\cite{Hill_PRB_72_045350_2005}.  The results for this appear in Table~\ref{tab:gate_times}.  As in Ref.~\onlinecite{Hill_quant-ph_0610059_2006}, we assume that single qubit rotations by an angle $\pi$ take 40~ns to perform as does the Hadamard gate.  We also assume that two-qubit rotations by $\pi$/4 take 1.96~ns if the coupling strength is given by $J_0=0.132\ \mu$eV, taken from the calculated unbiased exchange data\cite{Wellard_PRB_68_195209_2003}.  Actual time will decrease under the application of a $J$-gate bias\cite{Wellard_PRB_68_195209_2003, Wellard_J_Phys_Condes_Matter_16_5697_2004}, however, we assume a worse case scenario here.
\begin{table}[h]
\caption{CNOT gate times for various pulse implementation levels in the electron spin solid-state quantum computing architecture.}
\renewcommand{\arraystretch}{1.25}
\label{tab:gate_times}
\begin{center}
\begin{tabular}{cccc} \hline\hline
implementation & \multicolumn{3}{c}{gate times (ns)}\\
level & \quad single qubit \quad & \quad two-qubit \quad & \quad total \quad\\
\hline
0 & \quad 180\quad & \quad 3.92\quad & \quad 183.92\quad\\
1 & \quad 716\quad & \quad 35.28\quad & \quad 751.28\quad\\
2 & \quad 53256.80\quad & \quad 2544.08\quad & \quad 55800.88\quad\\
\hline\hline
\end{tabular}
\end{center}
\end{table}

As Table~\ref{tab:gate_times} demonstrates, operation time grows appreciably with concatenation.  Furthermore, Fig~\ref{fig:fidelity_exchange_100_direction} shows that the success of the robust CNOT gate is dependent on how accurately we can estimate the exchange coupling strength based on expectations of the fabrication process alone.  In such an uncharacterised system we have shown that a sensible choice can be made based upon the target separation, yielding $J_0$.  Large variations in the exchange interaction strength due to donor misplacement, and the additional time cost for multiple concatenation means composite rotations alone can not always guarantee a feasible, robust CNOT gate.  However, we will now show that composite pulses at the lowest level coupled with a systematic two-qubit interaction characterisation procedure allows for precise CNOT gate construction.

%
\section{The role of two-qubit Hamiltonian characterisation}
\label{sect:Charactersiation}
Using a combination of system indentification and composite rotations, we may construct a high fidelity robust CNOT gate.  Whilst many methods of system identification exist, we choose the procedure of Hamiltonian characterisation because it provides direct knowledge of the Hamiltonian (which we require) in an efficient manner.  This approach strikes a balance between the need for multiple concatenation and precision Hamiltonian characterisation, and may be particularly useful for systems whose Hamiltonian parameters require re-characterisation over time due to drift.

Recent work shows how characterisation of a two-qubit Hamiltonian can be achieved via entanglement mapping of the squared concurrence relation\cite{Cole_JPhysA_39_14649_2006,Devitt_Phys_Rev_A_73_052317_2006}.  The identification of the Hamiltonian coefficients amounts to determining the oscillation frequency of this entanglement function for different input states.  The only requirements are an accurately characterised Hadamard gate and measurement on both qubits.  An important result from the work in Ref.~\onlinecite{Cole_JPhysA_39_14649_2006}, is the fractional uncertainty in a frequency determination
\begin{equation}
\frac{\delta f}{f}\geq\frac{4}{N_t\sqrt{N_e}},
\label{eqn:frequency_fractional_uncertainty}
\end{equation}
where, $N_t$ is the number of discrete time points at which $N_e$ projective measurements are made.  An equivalent result can also be found in the earlier work of Huelga et al. in the context of Ramsey spectroscopy\cite{Huelga_PRA_79_3865_1997}.  To accurately determine the frequency, the time over which the system is observed, $t_{\rm ob}$, should be maximised, however this process is limited by the decoherence time of the system.  An accurate frequency determination is still possible in the presence of decoherence by allowing $t_{\rm ob}$ to be relatively large and performing two measurements at $N_t$ time points.  The uncertainty in the frequency can then be reduced by evolving the system for a suitably long time before measuring at two final time points. This process is repeated $N_e$ times to estimate the phase of the oscillation.  The total number of required measurements is then $N=2(N_t+N_e)$.  Characterising the system in this way results in the scaling of Eq.~\ref{eqn:frequency_fractional_uncertainty}.

To characterise the Heisenberg Hamiltonian with isotropic couplings requires determining the oscillation frequency of three different input states, meaning $N=6(N_t+N_e)$ total measurements are needed.  The fractional uncertainty in the characterised exchange coupling, $J_{\rm c}$, as a function of $N$ for a given $N_t$ is 
\begin{equation}
\frac{\delta J_{\rm c}}{J_{\rm c}}\equiv\frac{\delta f}{f}\geq\frac{4\sqrt{6}}{N_t\sqrt{N-6N_t}}.
\label{eqn:J_fractional_uncertainty}
\end{equation}
To illustrate the effect of composite rotations we consider a modest amount of characterisation by choosing $N_t=10$.  Increasing the number of time points results in higher precision characterisation.

In an uncharacterised system we assumed the coupling between donors, $J_0$, to be determined by the target donor separation.  Donor misplacement as a result of fabrication uncertainties lead to variations in the coupling strength, $J$, from the target $J_0$.  We have seen how the robust CNOT gate for an uncharacterised system performs in Fig.~\ref{fig:fidelity_exchange_100_direction}.  We now consider the performance of a robust CNOT gate in a characterised system.

Characterisation of the Hamiltonian can be performed to any level of precision at the expense of extra measurements, with the uncertainty given by Eq.~\ref{eqn:J_fractional_uncertainty}.  In a characterised system, the estimated coupling strength is set to the characterised coupling strength, $J_{\rm c}$ (with uncertainty bounds $\pm\delta J_{\rm c}$), rather than $J_0$.  The fractional error in this case is
\begin{equation}
\Delta_{\rm c}=\frac{J}{J_{\rm c}}-1,
\label{eqn:delta_definition_J_characterised}
\end{equation}
where in general the characterised coupling strength, $J_{\rm c}$, will be much closer to the true value of $J$, than the target value, $J_0$ is to $J$.  This means higher fidelity can be achieved using fewer levels of concatenation.

\begin{figure}[tb]
\centerline{\includegraphics{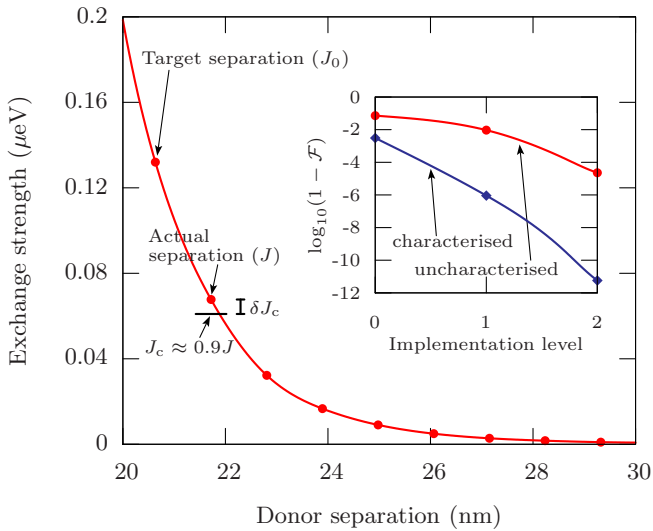}}
\caption{Exchange interaction strength as a function of donor separation along the [100] direction, showing a large variation in the coupling strength with donor misplacement (dots indicate actual site separations). For an uncharacterised system the coupling is set to the fabrication target $J_0$, with the actual placement giving coupling $J$.  The resulting CNOT error, ($1-{\cal F}$), for a one site deviation ($\Delta_0\approx -0.49$) from the target separation can be seen on the inset plot as a function of implementation level.  In the characterised system the coupling is set to $J_{\rm c}$.  The CNOT error for a system characterised to the 10\% level ($\delta J_{\rm c}/J_{\rm c}=0.1$), taking $J_{\rm c}\approx 0.9J$ ($\Delta_{\rm c}=0.1$), is shown as a function of implementation level inset also.  Note that all curves are included purely to guide the eye.}
\label{fig:fidelity_with_characterisation_100_direction}
\end{figure}

Given that the total gate time increases so sharply with increased concatenation, operating with a single application of the composite rotations is preferential.  For a one site deviation from the target separation, we show the resulting CNOT fidelity as a function of pulse implementation in a system characterised to the 10\% level ($\delta J_{\rm c}/J_{\rm c}=0.1$) in Fig.~\ref{fig:fidelity_with_characterisation_100_direction}.  Characterisation to this level would require at least 156 measurements assuming the previous parameters.  We take $J_{\rm c}\approx 0.9J$ to be the characterised value of the exchange coupling strength, as it corresponds to an extremal bound value.
The results in Fig.~\ref{fig:fidelity_with_characterisation_100_direction} demonstrate that it is possible to construct a very high fidelity CNOT gate using one level of robust pulsing, provided a suitable amount of characterisation is first performed.

\begin{figure}[tb]
\centerline{\includegraphics{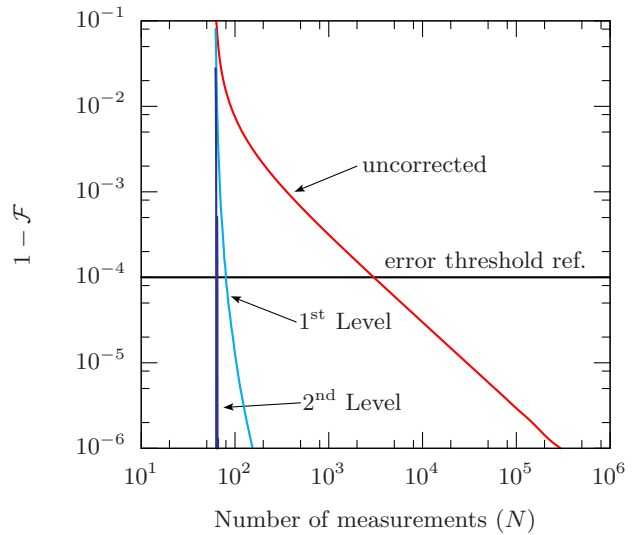}}
\caption{CNOT error, ($1-{\cal F}$), as a function of the total number of characterisation measurements required to achieve a given fidelity for various implementation levels.  The results demonstrate the usefulness of combining composite rotations with Hamiltonian characterisation when constructing a robust CNOT gate.  Threshold reference line at $10^{-4}$ error rate is shown.}
\label{fig:fidelity_vs_measurements}
\end{figure}

The total number of characterisation measurements needed to achieve a given fidelity can also be determined as a function of the implementation level.  These results appear in Fig.~\ref{fig:fidelity_vs_measurements}.
In reality the fidelity may be substantially higher than the results of Fig.~\ref{fig:fidelity_vs_measurements} indicate, as they provide a lower bound for the corresponding number of measurements.  These results show the clear benefit in using a single level of composite rotations and characterisation to construct a robust CNOT gate.  The improvements expected beyond this do not seem to warrant concatenation.

Any quantum computation proposal requires that many operations be performed within the dephasing time, $T_2$, of the system.  The $10^{-4}$ level is widely assumed to be the fault-tolerant threshold for both environmentally induced and systematic errors\cite{Gottesman_PhD_Thesis_1997}, however more rigorous bounds\cite{Aliferis_QIC_6_97_2006} recently calculated, suggest it could be closer to $10^{-5}$.  Figure~\ref{fig:fidelity_vs_measurements} shows that it is possible to construct a CNOT gate to this precision level in the presence of significant fabrication induced uncertainties, using either multiple concatenation of the composite rotations or a combination of the composite rotations and a modest level of characterisation.

\begin{figure}[tb]
\centerline{\includegraphics{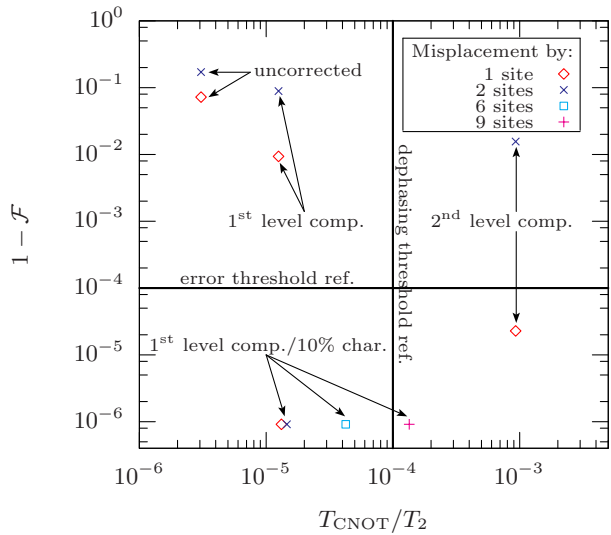}}
\caption{CNOT error, ($1-{\cal F}$), as a function of the total gate time for an unbiased, $J(V=0)$, system ($T_2=60\ \mbox{ms}$ assumed).  Results are shown for a range of separations in the [100] direction, larger than the targeted 20.634~nm separation.  We consider various CNOT gate constructions, namely an uncorrected CNOT, one constructed from both a single and two applications of composite rotations and finally a CNOT constructed using composite rotations in conjunction with characterisation to the 10\% level ($\delta J_{\rm c}/J_{\rm c}=0.1$) taking $J_{\rm c}\approx 0.9J$.  Only for this final method have more than two sites been included as for other methods results will clearly be worse.  Threshold reference lines at $10^{-4}$ error rates are shown.}
\label{fig:fidelity_vs_CNOT_time}
\end{figure}

Assuming the system has been characterised to a modest level beforehand, we now show that in order to remain below the threshold for environmentally induced errors also, the robust CNOT should be constructed using a single application of composite rotations and characterisation.  In Fig.~\ref{fig:fidelity_vs_CNOT_time}, these results are shown for a system with an unbiased $J$-gate, $J(V=0)$, based on the 60~ms dephasing time in isotopically pure $^{28}$Si at 7~K, and for characterisation to the 10\% level, again assuming the extremal bound value of $J_{\rm c}\approx 0.9J$.  
In a biased system, the exchange coupling is stronger.  Calculations suggest that for donors separated by $\sim 20\ \mbox{nm}$ in the [100] direction, a 1~V bias applied to the control gates can strengthen the coupling by over two orders of magnitude\cite{Wellard_PRB_68_195209_2003, Wellard_J_Phys_Condes_Matter_16_5697_2004}.  A robust CNOT gate comprising characterisation as described above could therefore operate at close to the $10^{-7}$ level for environmentally induced errors.  Performing additional measurements to characterise the system to the 1\% level would lower the systematic error level to well below $10^{-7}$ also, bringing it well within more rigorous threshold bounds\cite{Aliferis_QIC_6_97_2006}.

For systems whose Hamiltonian parameters are not well known due to fabrication uncertainties, or may drift over time, this is an important result, suggesting that operating the CNOT gate in this way can guarantee that the error rate remains below the fault-tolerant error threshold.  For the case of Si:P quantum computer architectures Fig.~\ref{fig:fidelity_vs_CNOT_time} suggests that this may be fabrication uncertainties within up to six sites of the target site, or $\sim 6.5\ \mbox{nm}$ in the unbiased case, however in the $J$-gate biased case this allowance may be much greater.  The trade-off for operating in this manner is the need for periodic re-characterisation, however the cost of this should be minimal as the number of required measurements is small.

%
%
\section{Conclusions}
The performance of a robust CNOT gate constructed using two-qubit composite rotations has been examined.  Multiple concatenation of the composite rotations results in a high fidelity CNOT gate provided the fractional uncertainty in $J$ lies within the correctable range.  Large variations in the exchange interaction coupling with donor separation means this is not always the case.  Furthermore, multiple concatenation of composite rotations requires long overall gate times with respect to the dechorence time of the system and results in gate operation which exceeds the current error threshold required for fault-tolerant quantum computation.  As an effective fix to this problem, we demonstrated how, in a system with large variations in the qubit coupling strength, a high fidelity CNOT gate which operates below this error threshold can be constructed from a single level of composite rotations in conjunction with Hamiltonian characterisation.

\section*{Acknowledgments}
The authors would like to thank Jared H. Cole for helpful discussions.  This work was supported by the Australian Research Council, the Australian Government and by the US National Security Agency (NSA), Advanced Research and Development Activity (ARDA), and the Army Research Office (ARO) under contract number W911NF-04-1-0290.  CDH is supported by EPSRC grant number EP/C012674/1.


\bibliography{composite_pulses_and_J}

\end{document}